\documentclass{aastex631}

\newcommand{\citepinprep}[1]{(#1 et al., \textcolor{blue}{in prep.})}
\newcommand{\citetinprep}[1]{#1 et al. (\textcolor{blue}{in prep.})}

\begin{document}

\title{Subhalos in Galaxy Clusters: Coherent Accretion and Internal Orbits}

\correspondingauthor{Chi Han}
\email{hanchi@umich.edu}
\author[0009-0003-1051-2624]{Chi Han}
\affiliation{Department of Physics, University of Michigan, Ann Arbor, MI 48109, USA}
\affiliation{Department of Astronomy, University of Michigan, Ann Arbor, MI 48109, USA}

\author[0000-0001-7690-2260]{Kuan Wang}
\affiliation{Department of Physics, University of Michigan, Ann Arbor, MI 48109, USA}
\affiliation{Leinweber Center for Theoretical Physics, University of Michigan, 450 Church St, Ann Arbor, MI 48109, USA}

\author[0000-0001-8868-0810]{Camille Avestruz}
\affiliation{Department of Physics, University of Michigan, Ann Arbor, MI 48109, USA}
\affiliation{Leinweber Center for Theoretical Physics, University of Michigan, 450 Church St, Ann Arbor, MI 48109, USA}

\author[0000-0003-3312-909X]{Dhayaa Anbajagane}
\affiliation{Department of Astronomy and Astrophysics, University of Chicago, Chicago, IL 60637, USA}
\affiliation{Kavli Institute for Cosmological Physics, University of Chicago, Chicago, IL 60637, USA}

\begin{abstract}
Subhalo dynamics in galaxy cluster host halos govern the observed distribution and properties of cluster member galaxies. 
We use the IllustrisTNG simulation to investigate the accretion and orbits of subhalos found in cluster-size halos.
We find that the median change in the major axis direction of cluster-size host halos is approximately $80$ degrees between $a\sim0.1$ and present-day.  
We identify coherent regions in the angular distribution of subhalo accretion, and $\sim68\%$ of accreted subhalos enter their host halo through $\sim38\%$ of the surface area at the virial radius.
The majority of galaxy clusters in the sample have $\sim2$ such coherent regions.
We further measure angular orbits of subhalos with respect to the host major axis and use a clustering algorithm to identify distinct orbit modes with varying oscillation timescales.
The orbit modes correlate with subhalo accretion conditions.
Subhalos in orbit modes with shorter oscillations tend to have lower peak masses and accretion directions somewhat more aligned with the major axis.
One orbit mode, exhibiting the least oscillatory behavior, largely consists of subhalos that accrete near the plane perpendicular to the host halo major axis. 
Our findings are consistent with expectations from inflow from major filament structures and internal dynamical friction: most subhalos accrete through coherent regions, and more massive subhalos experience fewer orbits after accretion. 
Our work offers a unique quantification of subhalo dynamics that can be connected to how the intracluster medium strips and quenches cluster galaxies.
\end{abstract}


\section{Introduction}
\label{sec:intro}

Structures form hierarchically in the concordance $\Lambda$CDM model of modern cosmology, supported by cosmic microwave background measurements and cosmology surveys of large scale structure \citep[e.g.,][]{Planck2018_cosmoparams,DES2018}.
In this model, dark matter halos form around density peaks in the initial fluctuation, and grow through mergers and accretion.
Galaxies reside in dark matter halos, which are the primary determining factor of galaxy formation and evolution \citep[e.g.,][]{white_rees78,blumenthal84}.
Over the past few decades, cosmological simulations have been extensively used to study the large-scale structure of the Universe \citep[see, e.g.,][]{Vogelsperger2020}. 
$N$-body simulations follow the gravitational interactions of dark matter \citep[e.g.,][]{Millennium,bolplanck2016,klypin_etal16}, and hydrodynamical simulations model various physical processes in galaxy formation and evolution \citep[e.g.,][]{Illustris,schaye2015_eagle,dubois2016}.
These simulations capture the distribution of cosmic voids, sheets, filaments, and nodes.
Galaxy clusters reside in the largest and most massive of the nodes.  

As the most massive of the gravitationally collapsed structures in the universe, galaxy clusters contain hundreds or thousands of galaxies, with masses above $10^{14}M_\odot$.
Galaxy clusters are also relatively late forming objects in our universe, whose population abundance and evolution is sensitive to dark energy \citep[e.g.,][]{allen2011,weinberg2013}.
Galaxy cluster observations range across the wavelength spectrum.
Observations probe the hot plasma that comprises the intracluster medium in the X-ray \citep[e.g.,][]{Chandra2000_Xray,liu2022erosita} and microwave \citep[e.g.,][]{WMAP_SZ,bleem2023spt500}, radio signatures from radio relics and active radio galaxies \citep[e.g.,][]{vanweeren2019review,lee2023}, and the cluster galaxy population in the optical and UV \citep[e.g.,][]{rykoff2016,maturi2019,aguena2021}.  

Cosmological simulations are an essential ingredient to our understanding of galaxy cluster astrophysics and cosmology \citep[e.g.,][]{kravtsov_borgani2012}.  In particular, simulations enable us to explicitly trace the evolution of galaxy clusters over cosmic time and connect their histories with observable signatures \citep[e.g.,][]{nelson2012,chen2019,green2020,deluca2021}.  One specific area of interest is how dynamics in galaxy clusters impact the observed distribution and properties of cluster member galaxies.  For example, the merger history of and subhalo dynamics within galaxy clusters can affect the properties of the brightest cluster galaxy \citep[e.g.,][]{contrerassantos2022} and star formation rates of cluster galaxies \citep[e.g.,][]{hough2023,lopes2024}.  Coherent accretion of galaxy groups into the galaxy cluster potential well can also leave imprints in the distribution and properties of cluster member galaxies \citep[e.g.,][]{roan2023}.  Underlying the cluster galaxy population is the subhalo distribution, which can be probed by gravitational lensing \citep[e.g.,][]{srivastava2023}.
 
Galaxy clusters have underlying halo shapes that are triaxial \citep[e.g.,][]{allgood2006}.  Observations have shown a morphological shape alignment of different tracers of galaxy cluster mass distributions, including X-ray, SZ, and lensing measurements \citep[e.g.,][]{donahue2016}.
The major axes of dark matter host halos typically align with major filament directions \citep[e.g.,][]{faltenbacher2005,Morinaga2020_orientation}, and halo shapes tend to be more elongated for faster accreting clusters \citep[e.g.,][]{chen2019,machado2021,lau2021,Anbajagane2022BarImp}.
Such alignments are also manifest in galaxies.
Multiple observational works have found that, in galaxy clusters, the orientation of the central galaxy is aligned with the distribution of satellite galaxies and the surrounding large-scale structure \citep[e.g.,][]{Yang2006,Huang2016,rodriguez2022,smith2023}.
Findings of recent works that studied hydrodynamical simulations \citep[e.g.,][]{Ragone-Figueroa2020,shi2021,rodriguez2023,Zhang_2023} suggest that the alignment between central and satellite galaxies originates from their alignment with dark matter halos and the cosmic web.

The accretion and orbits of subhalos found within cluster-size halos determine the observed distribution of galaxies in clusters and their properties. Theoretical studies of the phase-space of infalling subhalos indicate that satellites preferentially accrete along the major axis of a host dark matter halo \citep[e.g.][]{Wang2005_phase_space}. Observations have suggested that the satellite galaxies, which subhalos host, accrete along filaments \citep[e.g.,][]{wang2020}.  
Therefore, the direction through which most galaxy cluster subhalos have accreted should correspond to cosmic filaments that feed into the galaxy cluster and connect that galaxy cluster to the cosmic web. The subhalo orbits, both before and after accretion, reflect gravitational interactions, including tidal effects and dynamical friction. For example, observations have found distinct differences in how red and blue satellite galaxies align with tracers of the host halo major axis, with red galaxies showing stronger alignment with the major axis \citep[e.g.,][]{faltenbacher2007}.  We also expect the cluster galaxy distribution to vary depending on subhalo accretion.  These interactions have implications for galaxy formation in cluster environment,  galaxy-halo connection models in galaxy clusters, and cluster-based cosmology measurements based on galaxy measurements. 

In this work, we study cluster-size halos and their subhalos in the IllustrisTNG simulations.
We examine the evolution of host halo orientation, identify coherent regions of subhalo accretion onto host halos, and distinguish between subhalo orbit modes within hosts.
We organize this paper as follows.
In \autoref{sec:sim_method}, we introduce the simulation data and methods used in our analysis. In \autoref{sec:results}, we present our results. We draw conclusions and discuss our findings in \autoref{sec:discussion}.

\section{Simulation and Methods}
\label{sec:sim_method}

\subsection{IllustrisTNG simulations}
\label{sec:tng}

In this work, we analyze objects from the IllustrisTNG simulations \citep{marinacci2018_TNG,naiman2018_TNG,nelson2018a_TNG,pillepich2018b_TNG,springel2018_TNG,nelson2019a_TNG}.
IllustrisTNG is a suite of large volume, cosmological, gravo-magnetohydrodynamical simulations run with the \textsc{AREPO} code \citep{springel_2010}. The simulations adopt a cosmology consistent with \citet{planck2016}, with $\Omega_{\Lambda, 0} = 0.6911$, $\Omega_{m,0} = 0.3089$, $\Omega_{b,0} = 0.0486$, $\sigma_8 = 0.8159$, $n_s = 0.9667$, and $h = 0.6774$. For each run, 100 snapshots are stored, from early time to the present day, $z=0$. In this study, we use the TNG300-1 run, which has a box size of $205h^{-1}{\rm Mpc}$, a dark matter resolution of $4\times 10^7h^{-1}M_\odot$, and a baryon mass resolution of $7.6\times10^6h^{-1}M_\odot$.

In the IllustrisTNG simulations, halos are identified using the friends-of-friends (FoF) algorithm \citep[e.g.,][]{davis85_fof}, with a linking length of $b=0.2$.
Subhalos, which host individual galaxies, are identified with the \textsc{Subfind} algorithm \citep{subfind}.
The merger trees at the subhalo level are constructed with the \textsc{SubLink} algorithm \citep{rodriguez-gomez2015}.

\subsection{Halo sample and assembly history}
\label{sec:halos_mah}

For this analysis, we select cluster-size halos in the present-day snapshot of TNG300-1, with virial masses above $10^{14}M_\odot$.
This results in a sample of 376 halos.
For each halo in the sample, we construct its mass assembly history and record its mass and position as functions of time.
We construct the halo assembly histories with an algorithm similar to \citet{Wang2023_mergers}, using the package \textsc{TreeHacker} \citepinprep{Wang}
We briefly describe the procedure below and refer the interested reader to \citet{Wang2023_mergers} for more details.

For a halo in the present-day snapshot, we adopt the subhalo with the most massive history \citep{DeLucia_Blaizot07} identified by \textsc{SubLink} as the central subhalo\footnote{In the language of IllustrisTNG, the central subhalo refers to the smooth component of the host halo.}.
We construct the main branch history of the central subhalo by recursively following the first progenitor identified by \textsc{SubLink}, and identify the halos that host the subhalos along the main branch history, which form the main branch history at the halo level.

\subsection{Host halo major axis}
\label{sec:major_axis}

We use the direction of the host halo major axis to quantify the evolution of halo orientation.
The major axis calculation, along with a catalog of other halo properties, will be described in detail in \citetinprep{Anbajagane}, and we provide a summary of the algorithm below.

We calculate the major axis direction of a halo based on the gravitationally bound dark matter particles.
We perform the unbinding procedure by comparing the velocity of each particle relative to the FoF center-of-mass velocity against its escape velocity calculated assuming an NFW profile \citep{nfw97}.
We then compute the reduced mass-inertia tensor \citep{zemp11},
\begin{equation}\label{eq:2}
    \mathcal{M}_{ij} = \sum^N_{k=0}\frac{x_{k,i}\,x_{k,j}}{r^2_\mathrm{ell,k}}.
\end{equation}
Here $r^2_\mathrm{ell,k}$ is the elliptical radius of the $k$th particle,
\begin{equation}\label{eq:1}
    r_\mathrm{ell}^2 = \frac{\tilde{x}^2}{a^2} + \frac{\tilde{y}^2}{b^2} + \frac{\tilde{z}^2}{c^2},
\end{equation}
where $\tilde{x}$, $\tilde{y}$, $\tilde{z}$ are the coordinates along the eigenvectors of the halo, and $a$, $b$, and $c$ correspond to the major, intermediate, and minor axes respectively.
Note that with the exception of this subsection, $a$ denotes the scale factor of the simulation snapshot throughout this paper.
We perform the shape estimation iteratively, applying the selection criterion $r_\mathrm{ell}^2<R_\mathrm{vir}^2$ after each iteration, until the estimated axis ratios $s=c/a$ and $q=b/a$ are consistent within 1\% between two consecutive iterations.
$a$ is set to 1, so that the semi-major axis is always $R_\mathrm{vir}$.
The direction of the eigenvector that corresponds to the eigenvalue $a^2$ is the direction of the major axis.

\subsection{Host halo dynamical time definition}
\label{sec:tdyn}

When examining the dynamical processes in halos, we measure time in units of the local dynamical timescale.
The dynamical time is the time required to orbit across an equilibrium dynamical system.
We adopt the definition in \citet{Mo_bookOfGalaxy}:
\begin{equation}
    \tau_{dyn}(t) = \sqrt{\frac{3\pi}{16G\bar{\rho}(t)}},
\end{equation}
where $G$ is the gravitational constant, $\bar{\rho}(t)$ is the mean density of the system at the cosmic time $t$. Following the conventions of \citet{Jiang_vdB16} and \citet{wang2020_merger}, we measure the time between two epochs in units of dynamical times as
\begin{equation}
    T_{\text{dyn}}(t(a);t(a_{\rm ref})) = \int^{t(a)}_{t(a_{\rm ref})}\frac{dt}{\tau_{\text{dyn}}},
\end{equation}
where $a$ is the epoch of interest, $a_{\rm ref}$ is the epoch of reference, and $t(a)$ and $t(a_{\rm ref})$ are the corresponding cosmic times.

\subsection{Subhalo position and mass tracking}
\label{sec:subhalo_tracking}

For each halo in our sample, we identify all of its subhalos in the present-day snapshot, and construct their main branch histories at the subhalo level by recursively following the first progenitor link identified by \textsc{SubLink}.
Note that we do not perform additional cuts on the subhalo mass or peak mass.
For each subhalo, we define the time of accretion as the scale factor at which the center of mass of the subhalo first crosses the virial radius of its host halo.
We record the angular position of the subhalo with respect to the host halo center at the time of accretion.
We track the mass evolution of each subhalo, and find the peak mass, which is the maximum subhalo mass in history.
Approximately 85\% of all subhalos in present-day clusters have evolution histories complete back to at least 4 dynamical times ago.
We describe the angular orbit of a subhalo in terms of its angular separation from the host halo major axis as a function of time.
Specifically, the angular separation refers to the angle between the vector pointing from the host halo center to the subhalo center, and the vector of the host halo major axis.

\section{Results}
\label{sec:results}

\subsection{Evolution of halo orientation}
\label{sec:results:orientation}

\begin{figure}
    \centering
    \includegraphics[width=8.5cm]{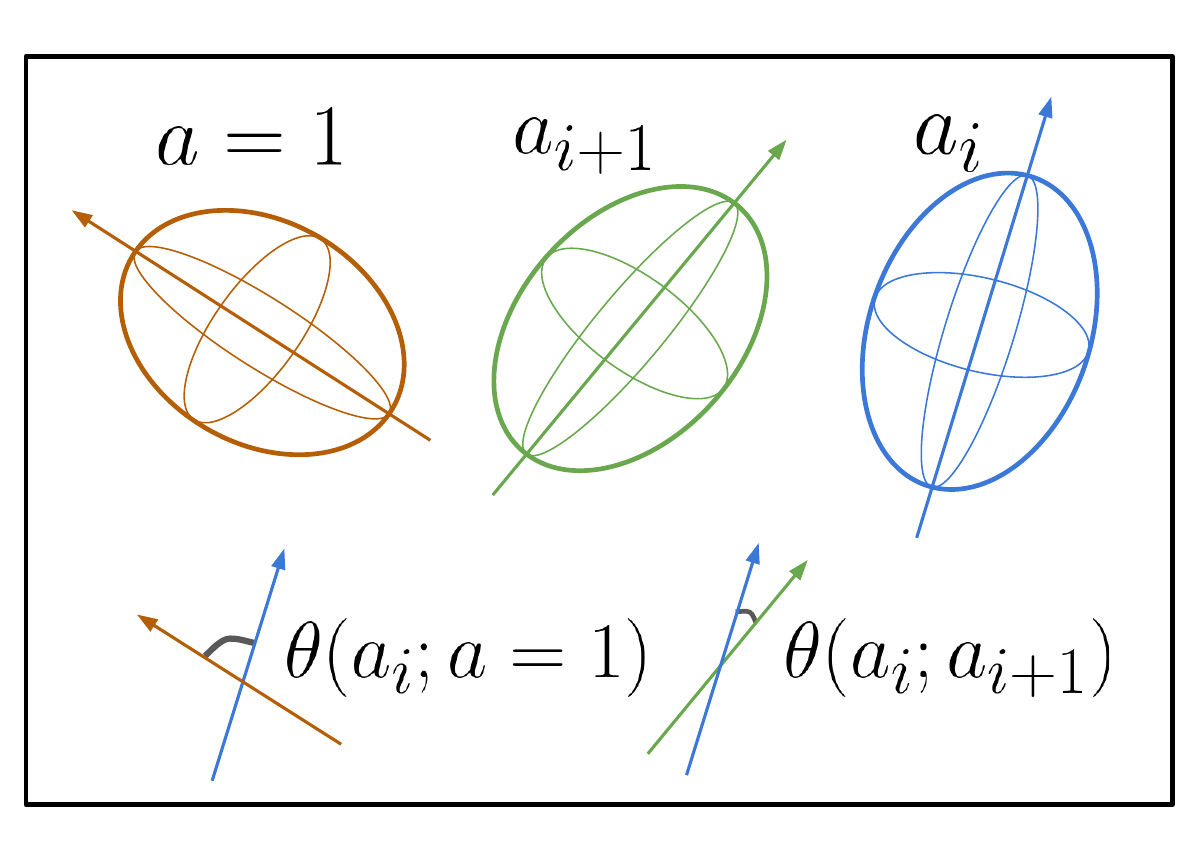}
    \caption{\textbf{Illustration of the change in halo orientation.}
    The vectors represent the major axis directions of the halo at different times.
    $\theta(a_i; a=1)$ is the angle between the directions of the major axis at snapshot $a_i$ and the major axis at the present day, whereas $\theta(a_i; a=1)$ represents the change in major axis direction between adjacent snapshots $a_{i}$ and $a_{i+1}$.}\label{fig:cartoon}
\end{figure}

We measure the orientation of a halo with the direction of its major axis, and track the evolution of the orientation throughout the assembly history of each halo.
For each halo, we choose the direction of its major axis at the present-day snapshot as the orientation of reference.
We then measure the angle $\theta(a_i;a=1)$ between the reference orientation and the orientation at each snapshot $a_i$.
The evolution of $\theta(a_i;a=1)$ over time can be used as a measure of the change in halo orientation during its evolution.
We note that $\theta(a_i;a=1)$ represents only one of two degrees of freedom in the angular space, and does not incorporate all the information of the orientation evolution.
We also measure the angle between adjacent snapshots, $\theta(a_i;a_{i+1})$, as a function of time\footnote{Note that the major axis has two opposite directions, and in our measurements, we choose the direction at each time step so that the change between adjacent snapshots is less than 90 degrees at all times, in order to avoid artificial sudden flips in the orientation.}.
We illustrate $\theta(a_i;a=1)$ and $\theta(a_i;a_{i+1})$ in \autoref{fig:cartoon}.

\begin{figure}
    \centering
    \includegraphics[width=0.5\linewidth]{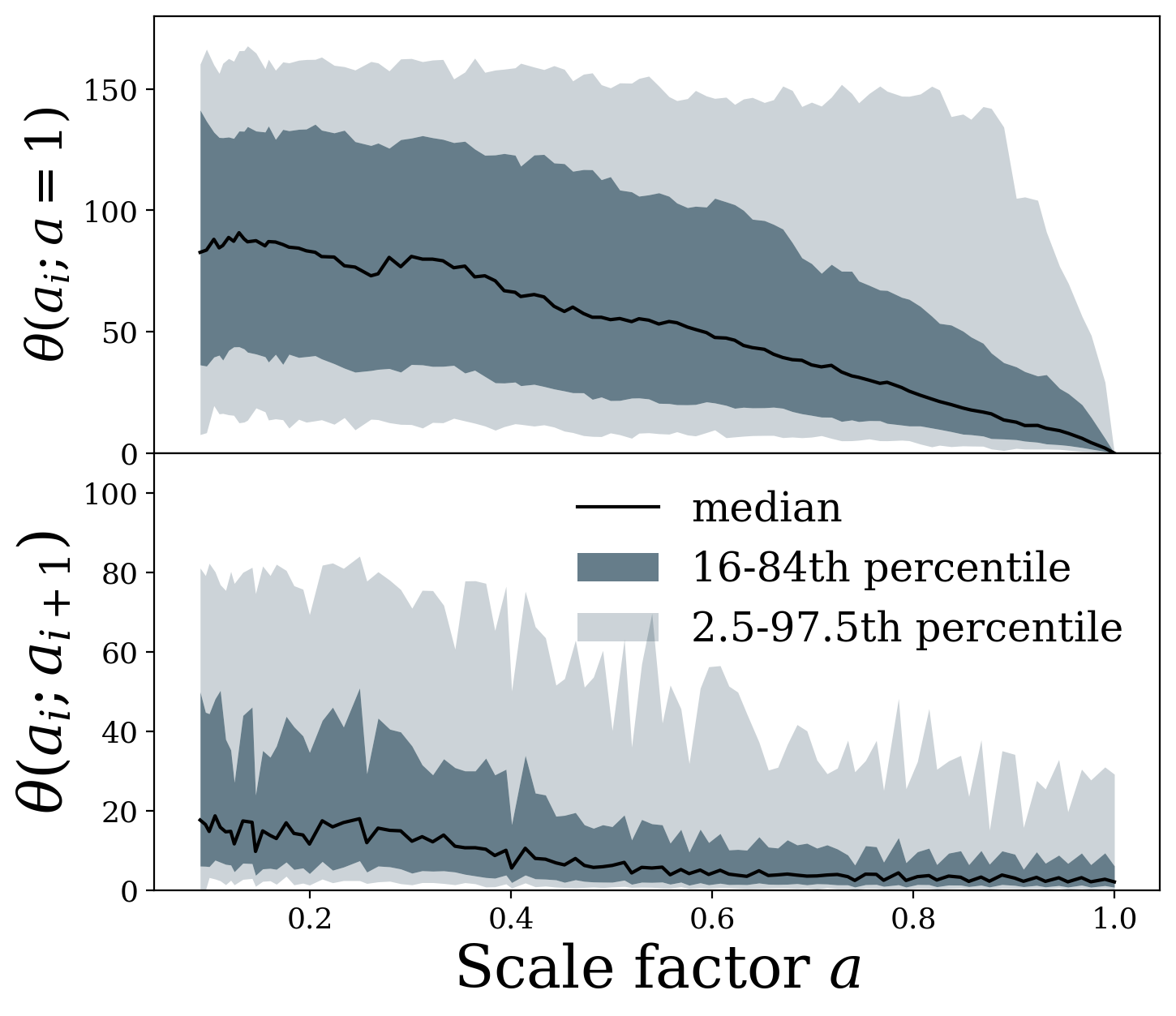}
    \caption{\textbf{Evolution of halo major-axis orientation.}
    In the top panel, the solid line shows the median orientation evolution of the halo sample with respect to the present-day, in terms of $\theta(a_i;a=1)$, as defined in \autoref{fig:cartoon}.
    The shaded regions show the 16-84th and 2.5-97.5th percentile ranges, as labeled in the figure.
    The bottom panel is similar, but shows the step-wise evolution of the halo orientation, $\theta(a_i;a_{i+1})$, for the same sample. 
    The halo population exhibits a moderate amount of change in orientation over cosmic time, and the change between simulation snapshots becomes slower at later times.}\label{fig:orientation}

\end{figure}

In the top panel of \autoref{fig:orientation}, we show the median evolution of $\theta(a_i;a=1)$, along with the 16-84th and 2.5-97.5th percentile ranges, for all the cluster-size halos in our sample.
We find a moderate amount of change in halo orientation over cosmic time, and the distribution of the angle shifts to larger values towards earlier times, which can be naively expected with angular momentum gained during accretion.
The halo population experiences a median change of $82.7^\circ$ since $a=0.09$, which is slightly below $90^\circ$.
From the percentile ranges, we find that $\theta(a=0.09;a=1)$ is broadly distributed between $0^\circ$ and $180^\circ$, which suggests that the information of early orientation is largely erased during the cosmic evolution.

We present the step-wise change in orientation, $\theta(a_i;a_{i+1})$, in the bottom panel of \autoref{fig:orientation}. 
We find that the majority of the halo sample experiences only mild changes in orientation between adjacent snapshots.
$\theta(a_i;a_{i+1})$ decreases with time in general: the median value is approximately $17.9^\circ$ at $a=0.09$, and $2.1^\circ$ at the present day.

\subsection{Coherent regions of subhalo accretion}
\label{sec:coherent}

As is described in \autoref{sec:subhalo_tracking}, we identify all the subhalos for each halo in our sample, and examine the distribution of their angular positions with respect to the center of the host at the time of accretion.  A subhalo time of accretion corresponds to when it enters the host halo virial radius.
Note that the angular positions used in this analysis are defined in the fixed coordinate system of the simulation, and are not dependent on the host major axis.
This procedure yields an angular map of accretion directions for each halo.
We use a spherical kernel density estimator (KDE)\footnote{We use the implementation described by \url{https://zenodo.org/records/4041472}\citep{spherical_kde}, which is available at \url{https://github.com/williamjameshandley/spherical_kde}} with a von Mises–Fisher kernel to create a continuous density function on the sphere.
We choose a kernel bandwidth of 0.25rad, which corresponds to a solid angle of approximately 820deg$^2$.
We note that this procedure does not discern structures in the angular distribution smaller than this scale.
However, our qualitative conclusions are not dependent on this choice.

\begin{figure}
    \centering
    \includegraphics[width=0.5\linewidth]{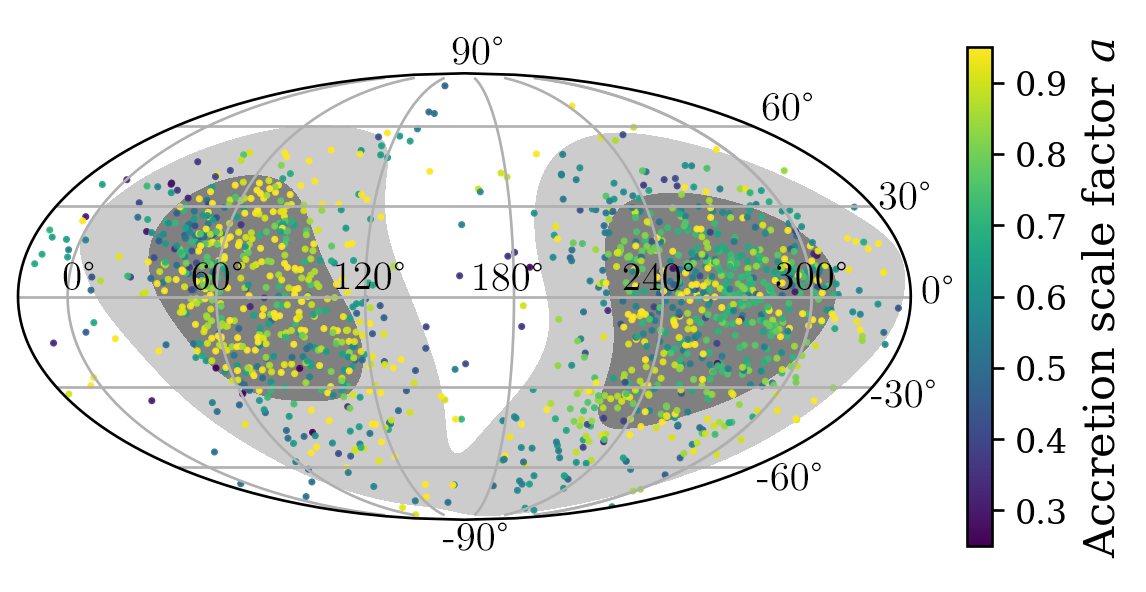}
    \caption{\textbf{Example KDE map for an individual host halo, illustrating two coherent regions in the angular distribution of subhalo accretion.}
    Each scatter point corresponds to a subhalo, and shows the angular position of accretion onto the host halo, color coded by the time of accretion.
    The shaded regions indicate regions of high number densities.
    The dark grey regions contain 68\% (1$\sigma$) of the subhalo accretion locations; the light grey region contains an additional 27\% (1-2$\sigma$) of subhalos accretion locations.
    The last 5\% percent of subhalos are accreted through the remaining region.}
    \label{fig:kde_example43}
\end{figure}

\autoref{fig:kde_example43} is an example angular map for a host halo randomly selected from the sample\footnote{Halo with GroupID 43 in the present-day snapshot.}.
This halo has a virial mass of approximately $4\times10^{14}h^{-1}M_\odot$, and we find 1404 subhalos in its virial boundary.
The orientation of this halo has changed by an angle of $35^{\circ}$ compared to $a=0.08$.
We show subhalo entry positions as scatter points and color code by their time of accretion.  We observe regions with high number densities in the distribution, showing that different subhalos tend to accrete onto the host halo along coherent directions.
We show the $1\sigma$ and $2\sigma$ contours of highest densities as shaded regions, where the $1\sigma$ region contains 68\% of the subhalos, and the $2\sigma$ region contains 95\%.
We hypothesize that these dense regions are likely physically connected to the filamentary structures in the cosmic web that feed the host halo, which we plan to further explore in future work.

\begin{figure}
    \centering
    \includegraphics[width=0.5\linewidth]{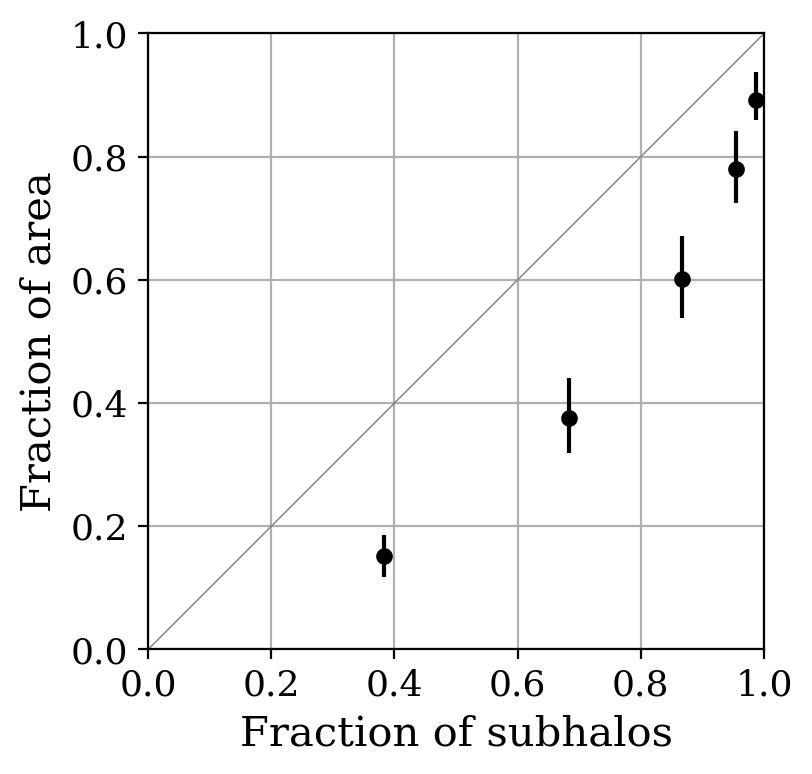}
    \caption{\textbf{Cumulative fraction of area as a function of fraction of enclosed subhalo accretion directions.}
    Fractional area on the celestial sphere of the host halo at the virial radius, through which a given fraction of subhalos accreted, as identified through a clustering algorithm illustrated in Figure~\ref{fig:kde_example43}.
    The data points correspond to the 0.5, 1.0, 1.5, 2.0, and 2.5$\sigma$ levels of subhalo fractions.
    The figure shows the median values for the halo sample, and error bars correspond to the 16-84th percentile range. 
    The majority of subhalos are accreted through compact regions.}
    \label{fig:coherent_frac}
\end{figure}

For all the cluster-size halos in our sample, we calculate the fraction of area in the 0.5, 1.0, 1.5, 2.0, and 2.5$\sigma$ regions (i.e., the regions that contain 38\%, 68\%, 87\%, 95\%, and 99\% of the subhalos).
In \autoref{fig:coherent_frac}, we show the fraction of area as a function of the fraction of subhalos to quantify coherent areas of subhalo accretion.
The error bars show the 16-84th percentile range for the halo sample around the median.
We find that approximately 68\% of the subhalos are accreted through $37\%$ of the total area, and 78\% of the area contains 95\% of the subhalos.

\begin{figure}
    \centering
    \includegraphics[width=0.5\linewidth]{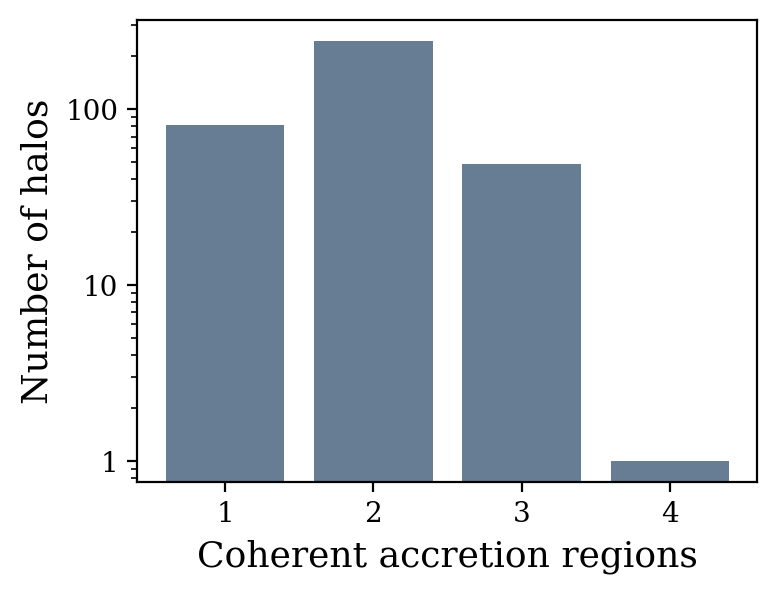}
    \caption{\textbf{Histogram of cluster-size halos with different numbers of coherent accretion regions.} Most galaxy cluster-size halos have 2 regions of coherent subhalo accretion, identified through the number of disjoint $1\sigma$ regions on their accretion direction maps, similar to Figure~\ref{fig:kde_example43}.
    Fewer halos have 1 or 3 disjoint regions; we find only one halo with 4 distinct accretion regions, and none with 0.
    Note that the $y$-axis is in logarithmic scale.}
    \label{fig:num-coherent-regions}
\end{figure}

We count the number of disjoint $1\sigma$ regions for each halo, and show a histogram in \autoref{fig:num-coherent-regions}.
We find that the majority of halos have 2 distinct 1$\sigma$ regions.
Most of the rest of the halos have 1 or 3 disjoint regions, with 1 region somewhat preferred over 3 regions.
We only find one halo with 4 accretion regions, and we do not find halos with more disjoint regions.

\subsection{Relative Position of Subhalos to the Major Axis}
\label{sec:rela_pos}

We now examine the internal orbits of subhalos, in terms of their angular separation from the host major axis.
In order to sufficiently track the internal orbits, we only include the subhalos accreted no less than 4 dynamical times ago (that is, before $a=0.47$) in this analysis.
This may introduce a survivor bias against subhalos that are disrupted within a shorter period of time of accretion, but otherwise we do not expect subhalos accreted at later times to have fundamentally different orbital behavior.
We perform this analysis with no cut on subhalo peak mass, however, we have tested that we get similar qualitative results when excluding subhalos with the 16\% and 50\% lowest peak masses.
Out of the two possibilities, we choose the vector direction of the initial halo major axis to ensure that the angle of accretion is between 0$^\circ$ and $90^\circ$.

Since the different orbits have varied time samplings, we first perform a linear interpolation with intervals of 0.02 dynamical times.
We use the k-means clustering algorithm to classify orbits into modes.
Due to the high-dimensional and bounded nature of these orbits, it is challenging to directly determine the optimal number of clusters to use.
We therefore run a principal component analysis (PCA) on the orbits.
We apply the Silhouette criterion to the first 10 principal components, and determine the optimal number of clusters for the k-means clustering algorithm to be 7.

\begin{figure}
    \centering
    \includegraphics[width=0.5\linewidth]{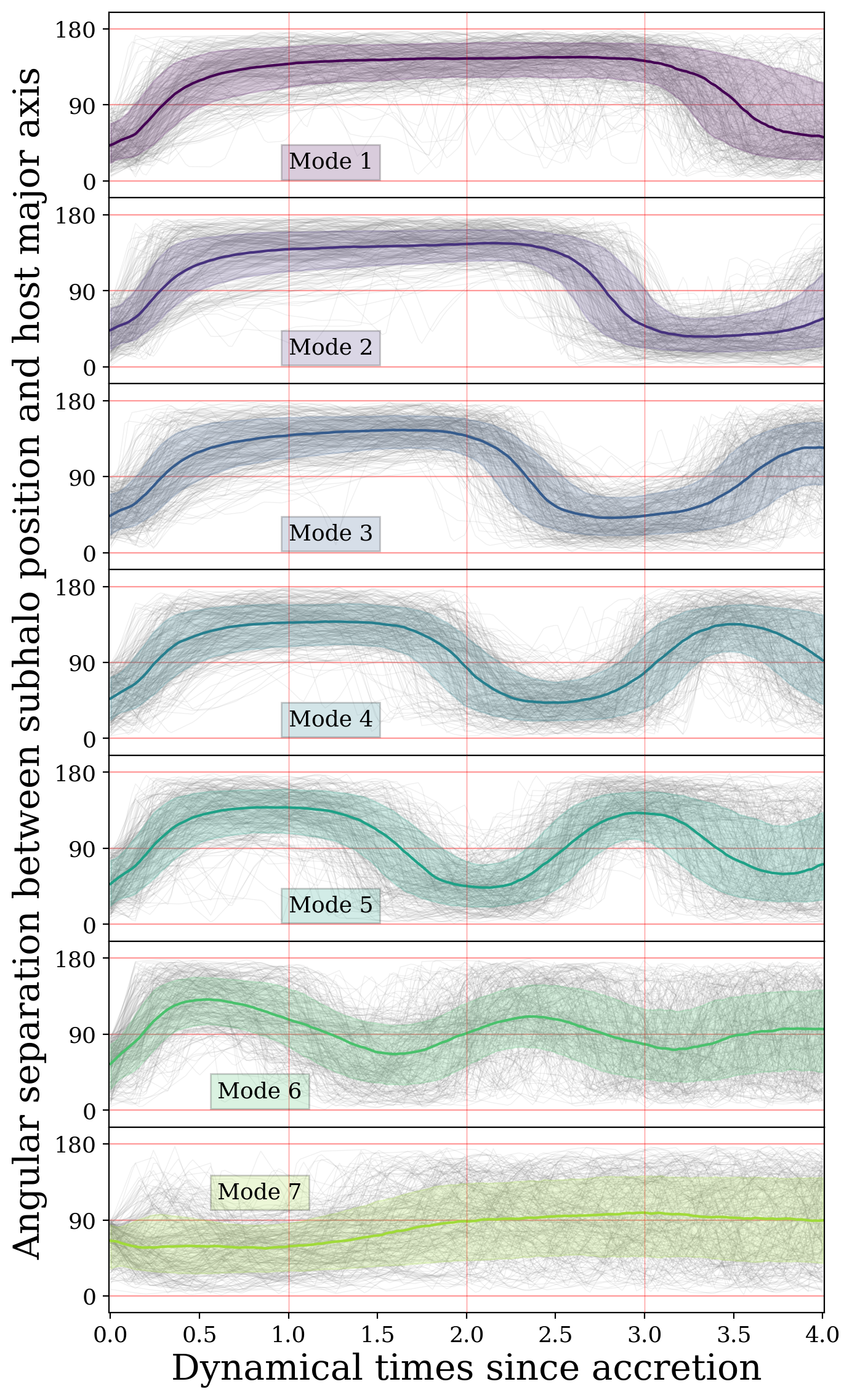}
    \caption{\textbf{Subhalo orbit modes classified by clustering algorithm.}
    In this figure we analyze the angular orbits of subhalos in their respective host halos.
    Time is measured in units of dynamical times, relative to the time of initial accretion.
    We find 7 clustered modes of orbits.
    We sort the modes by oscillation length over which subhalos maintain a relatively constant angular separation from their host halos.
    In each panel, the solid line indicates the median mode behavior, and the shaded area indicates the 16-84th percentile range.  
    We show individual orbits in thin grey lines.
}
    \label{fig:clusteredOrbit}
\end{figure}

We show the 7 orbit modes in \autoref{fig:clusteredOrbit}, where we plot individual orbits as thin grey lines in the background.  The thick colored line and band indicate the median and 16-84th percentile range for each mode.
We observe that for modes 1 to 6, subhalos first increase past $90^\circ$ within approximately 0.5 dynamical times and stay on an orbit of relatively stable angular separation from the major axis for different periods of time, before starting to oscillate around $90^\circ$.
The modes are sorted so that the time of the near-constant orbit decreases from mode 1 to mode 6.
Below we look into some factors that are potentially connected to the different mode behavior.

\begin{figure}
    \centering
    \includegraphics[width=0.5\linewidth]{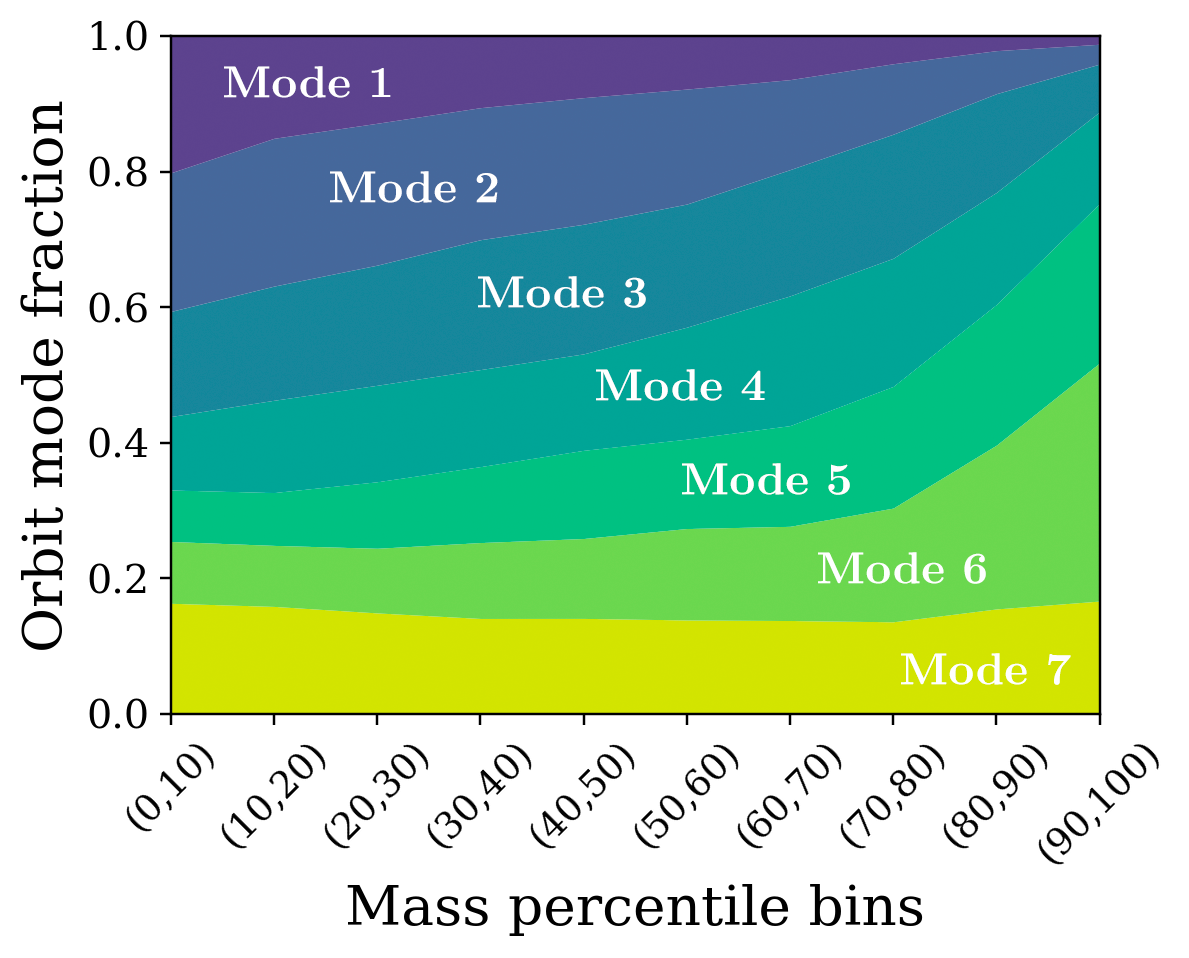}
    \caption{\textbf{Orbit mode fraction in different bins of subhalo peak mass.} 
    We bin the subhalos used in the mode classification by peak mass, and count the relative fraction of subhalos in each orbit mode within each percentile bin.
    From modes 1 to 6, we observe a trend that subhalos with higher peak masses dominate the lower orbit modes, corresponding to shorter oscillation times.  Mode 7 shows little to no peak mass preference.}
    \label{fig:mass_bin}
\end{figure}

In \autoref{fig:mass_bin}, we show the relative fractions of the orbit modes within each percentile bin of subhalo peak mass.
For modes 1-6, we find that lower peak masses favor the orbits with longer near-constant periods, whereas higher peak masses favor the orbits with shorter near-constant periods. 
This is consistent with our expectation that
more massive subhalos tend to experience stronger dynamical friction, lose more energy, orbit closer to the central region of the host, and exhibit shorter dynamical timescales \citep[e.g.,][]{jiang2015,vdb16}.
More recently, \citet{Wang2023_mergers}, led by one of the coauthors, looked into halo mergers in IllustrisTNG, and found that subhalos with smaller masses relative to their host halos have larger pericentric and apocentric distances, and reach these characteristic points at later times.
Mode 7, however, shows little dependence on the subhalo peak mass.
We argue that mode 7 subhalos are likely subhalos that orbit near the plane perpendicular to the major axis, which results in more chaotic and oscillatory angular evolution around $90^\circ$, consistent with what we observe in the bottom panel of \autoref{fig:clusteredOrbit}.

\begin{figure}
    \centering
    \includegraphics[width=0.5\linewidth]{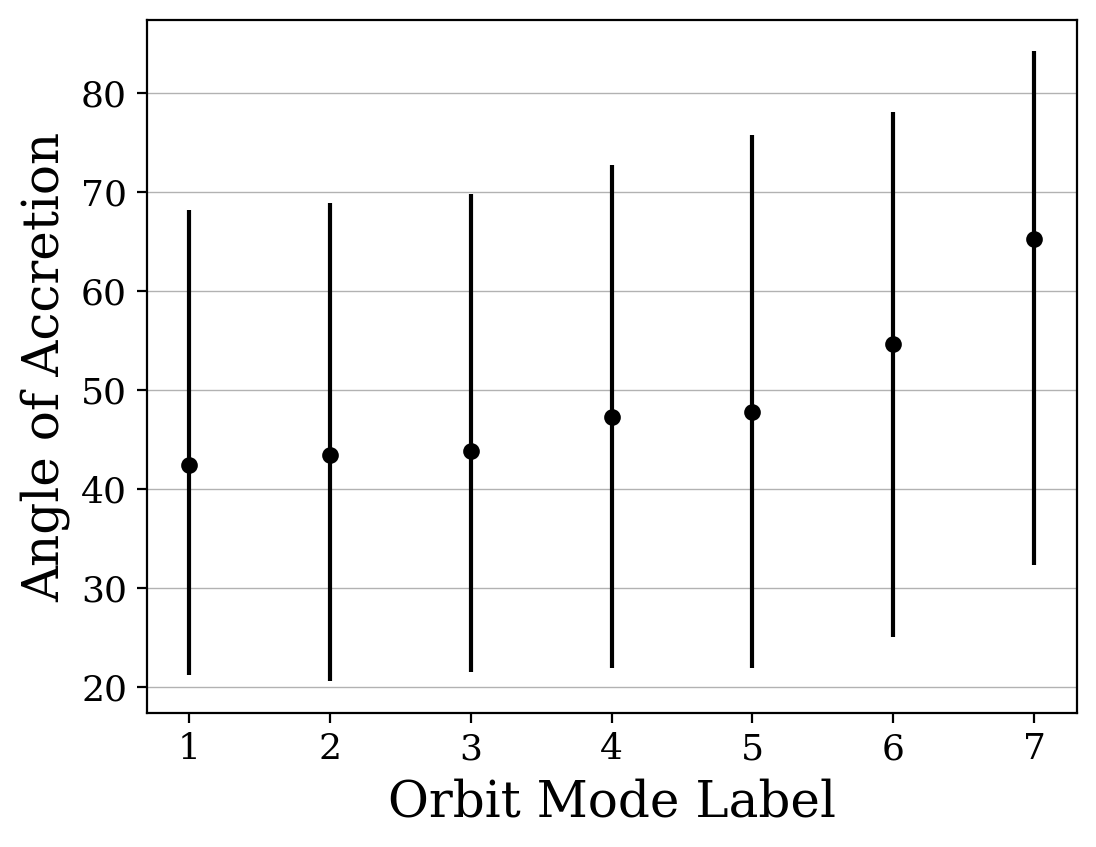}
    \caption{\textbf{Median angle of accretion with respect to host major axis of each orbit mode.}
    For each orbit mode, we measure the median angular separation between the host major axis and the subhalo position at the time of accretion.
    Error bars show the 16-84th percentile spread.
    We observe a weak but systematic trend that subhalos with longer oscillatory timescales typically accrete further from the major axis direction.}

    \label{fig:accretion_direction}
\end{figure}

In \autoref{fig:accretion_direction}, we measure the median and 16-84th percentile range of the angular position of subhalos at the time of accretion for each mode.
We find a weak but systematic trend for modes 1-6, where subhalos that accrete further away from the major axis tend to have shorter near-constant orbits.
The distribution of accretion angles of mode 7 subhalos is closer to $90^\circ$ than any other mode, which confirms our previous explanation of the behavior of this mode.

In \autoref{fig:mass_bin} and \autoref{fig:accretion_direction}, we find higher peak masses and accretion directions more perpendicular to the host major axis for the subhalos that have shorter dynamical timescales and experience more orbits.
These factors lead subhalos to approach central regions of the host halo faster and lose mass and energy more rapidly.
We notice in \autoref{fig:clusteredOrbit} that the difference in timescales is most pronounced during the first orbit, which indicates that the most significant differences between the modes occurs early in the dynamical evolution, in agreement with the findings of \citet{vdb16}, that the effect of tidal stripping enhances the mass loss rate and cause the subhalos to become less susceptible to dynamical friction in subsequent passages.
While the orbit behavior is statistically connected to these infall conditions, the evolution of individual subhalos is affected by many factors, such as the detailed mass distribution of the host halo, encounters with other subhalos, etc., and show significant stochasticity.

\section{Conclusions and Discussion}
\label{sec:discussion}

In this work, we study cluster-size halos and their subhalos in IllustrisTNG.
We examine the evolution of halo orientation in terms of the host halo major axis, the angular coherent accretion of subhalos, and subhalo orbits inside host halos.
These investigations lay the groundwork for connecting the orbits to galaxy properties and cluster gas properties in simulations.
We leave a focused analysis of baryonic properties to a follow-up paper to further the understanding of galaxy formation and evolution as cluster members, and to identify potential observational signatures of orbit and merger histories.

Below we summarize our findings:
\begin{itemize}
    \item Between early times ($a\sim0.1$) and the present day, cluster-size halos experience a median change of orientation slightly below 90 degrees.
    The step-wise change of orientation between adjacent snapshots is relatively mild and decreases with time.
    \item There are coherent regions in the angular distribution of the accretion direction of subhalos with respect to the center of the host. We conjecture that these regions correspond to filamentary structures connected with the host halo.
    \item Approximately 68\% (95\%) of the subhalos are accreted through 38\% (78\%) of the surface area.
    \item A vast majority of halos show one, two, or three disjoint $1\sigma$ regions in the angular distribution, where two is the most common number of regions.
    \item We find several distinct modes of subhalo orbits in host halos, with different oscillation timescales.
    \item We find that subhalos that have longer oscillation timescales tend to be accreted closer to the major axis, and have lower peak masses.
\end{itemize}

In this work, we have studied the orientation of cluster-size halos.
Halo orientation carries information of the halo formation history \citep[e.g.,][]{patiri2006}, and has implications for galaxy clustering and galaxy lensing measurements \citep[e.g.,][]{marrone2012}.
We have also looked into the accretion of subhalos onto these halos, which reflects the hierarchical assembly process in the large-scale structure, and sets the initial conditions for satellite galaxy dynamics.
Additionally, halo orientation and subhalo accretion are closely connected physically to each other through the surrounding cosmic web.
\citet{Wang2005_phase_space}, for example, showed that more massive subhalos tend to accrete along major axes and from the major filaments.
More recently, \citet{Morinaga2020_orientation} studied the impact of filamentary accretion of subhalos on halo orientation, and showed that halos with strong filamentary accretion have major axes are preferentially aligned with the filaments.

A potential avenue for future study is to compare the host halo major axis evolution, the accretion directions of subhalos, and the cosmic filaments constructed with cosmic web finding algorithms.
Different filaments may dominate halo mass assembly at different epochs, and another possibility is to study the time evolution of the coherent accretion regions of subhalos, and possibly also the motion and evolution of filaments themselves.
The major axis of a halo typically aligns with cosmic filaments \citep[e.g.,][]{faltenbacher2005,Morinaga2020_orientation}, and subhalos that enter the halo at a larger angle with respect to the major axis (mode 7 in \autoref{fig:clusteredOrbit}) are more likely accreted from voids and sheets.
It might be interesting to explore the different initial properties and orbits for subhalos accreted from different cosmic web environments in the future.

The initial accretion and subsequent orbits of subhalos are also naturally connected.
For example, \citet{gonzales2016} found that for cluster-size halos, subhalos accreted from filaments show somewhat longer lifetimes than subhalos accreted from other directions. 
They argued that massive hosts are connected to strong filaments with higher velocity coherence and density, which are possibly able to shield subhalos from the effect of dynamical friction during first infall.
The interplay of the accretion and the orbit shape the spatial and property distribution of satellite galaxies.
For instance, \citet{karp2023_quenching} recently found that satellite galaxies along the host halo major axis tend to have been accreted earlier and are hosted by subhalos with higher peak halo masses.
They argued that this correlations leads to anisotropic quenching of satellite galaxies.

We have found different modes of subhalo orbits and investigated some possible factors that affect the orbits.
\citet{aung2021}, among other works, showed that infalling and orbiting satellite galaxies in clusters belong to kinematically distinct populations.
An interesting future step is to apply our analysis pipeline to the two populations separately, and study whether they show systematically different orbital behavior, and to what extent this distinction can explain the different modes that we have found.

\section*{Acknowledgements}

We thank Gus Evrard, Farhan Hasan, Johannes Lange, Phil Mansfield, and Daisuke Nagai for useful discussions and feedback on early stages of this work.

CH acknowledges support from the Summer Undergraduate Research Fellowship from the Department of Physics at the University of Michigan.
KW and CA acknowledge support from the Leinweber Foundation at the University of Michigan.
CA acknowledges support from DOE grant DE-SC009193. 
DA is supported by NSF grant No. 2108168.

This research made use of Python, along with many community-developed or maintained software packages, including
IPython \citep{ipython},
Jupyter (\href{jupyter.org}{jupyter.org}),
Matplotlib \citep{matplotlib},
NumPy \citep{numpy},
SciPy \citep{scipy},
Pandas \citep{pandas}
and Astropy \citep{astropy, astropy1, astropy2}.
This research made use of NASA's Astrophysics Data System for bibliographic information.

\bibliography{main}{}
\bibliographystyle{aasjournal}



\end{document}